\documentclass[preprint,12pt]{elsarticle}

\usepackage{amssymb}
\usepackage{vdrcmac}
\usepackage{ccaption}
\usepackage{boxedminipage}
\usepackage{color}
\usepackage{comment}

\newcounter{bla}

\journal{Computer Physics Communications}

\begin{document}

\begin{frontmatter}



\title{MGOS: A Library for Molecular Geometry and Its Operating System}


% \author[a]{First Author\corref{author}}
% \author[a,b]{Second Author}
% \author[b]{Third Author}

% \cortext[author] {Corresponding author.\\\textit{E-mail address:} firstAuthor@somewhere.edu}
% \address[a]{First Address}
% \address[b]{Second Address}

\author[1,2]{Deok-Soo Kim\corref{author}}
\author[1]{Joonghyun Ryu}
\author[1]{Youngsong Cho}
\author[2]{Mokwon Lee}
\author[2]{Jehyun Cha}
\author[2]{Chanyoung Song}
%\author[2]{Jae-Kwan Kim}
\author[3]{Sangwha Kim}
\author[4]{Roman A Laskowski}
\author[5]{Kokichi Sugihara}
\author[6]{Jong Bhak}
\author[7]{Seong Eon Ryu}

\cortext[author] {Corresponding author.\\\textit{E-mail address:} dskim@hanyang.ac.kr;\\
J. Ryu and Y. Cho have equally contributed.}
\address[1]{Vorononi Diagram Research Center, Hanyang University, Korea}
\address[2]{School of Mechanical Engineering, Hanyang University, Korea}
\address[3]{College of Veterinary Medicine, Seoul National University, Korea}
\address[4]{European Bioinformatics Institute, Wellcome Trust Genome Campus, UK}
\address[5]{Meiji Institute for Advanced Study of Mathematical Sciences, Japan}
\address[6]{Department of BioMedical Engineering, UNIST, Korea}
\address[7]{Department of Bioengineering, Hanyang University, Korea}

\begin{abstract}
The geometry of atomic arrangement underpins the structural understanding of molecules in many fields.
However, no general framework of mathematical/computational theory for the geometry of atomic arrangement exists.
Here we present ``Molecular Geometry (MG)'' as a theoretical framework accompanied by ``MG Operating System (MGOS)'' which consists of callable functions implementing the MG theory.
MG allows researchers to model complicated molecular structure problems in terms of elementary yet standard notions of volume, area, etc. and MGOS frees them from the hard and tedious task of developing/implementing geometric algorithms so that they can focus more on their primary research issues.
MG facilitates simpler modeling of molecular structure problems; MGOS functions can be conveniently embedded in application programs for the efficient and accurate solution of geometric queries involving atomic arrangements.
%
% MG and MGOS facilitate simpler modeling and efficient, accurate, and convenient solution of molecular structure problems.
%
The use of MGOS in problems involving spherical entities is akin
to the use of math libraries in general purpose programming languages
in science and engineering.
\end{abstract}

\begin{keyword}
Atomic arrangement \sep 
Structural biology \sep
Material design \sep
Voronoi diagram \sep
Computational geometry \sep
Computational science
\end{keyword}

\end{frontmatter}

{\bf PROGRAM SUMMARY}

\begin{small}
\noindent
{\em Program Title: Molecular Geometry Operating System (MGOS)}                                          \\
{\em Licensing provisions: CC By 3.0}                                   \\
% {\em Licensing provisions(please choose one): 
% CC0 1.0/CC By 4.0/
% MIT/Apache-2.0/BSD 3-clause/BSD 2-clause
% /GPLv3/GPLv2/LGPL/
% CC BY NC 3.0/MPL-2.0 }                                   \\
{\em Programming language: C++}                                   \\
{\em Supplementary material: 
(1) Supplementary Video 1,
(2) Supplementary Video 2,
(3) Supplementary Video 3,
(4) Supplementary Video 4,
(5) MGOS manual, and
(6) 300 test PDB structure files}                                 \\
  % Fill in if necessary, otherwise leave out.
        \\
{\em Journal reference of previous version: N/A}                  \\
  %Only required for a New Version summary, otherwise leave out.
{\em Does the new version supersede the previous version?: N/A}   \\
  %Only required for a New Version summary, otherwise leave out.
{\em Reasons for the new version: N/A}\\
  %Only required for a New Version summary, otherwise leave out.
{\em Summary of revisions: N/A}*\\
  %Only required for a New Version summary, otherwise leave out.
        \\
{\em Nature of problem:}
\\
  %Describe the nature of the problem here. \\
For both organic and inorganic molecules, structure determines molecular function and molecular structure is highly correlated with molecular shape or geometry. 
Hence, many studies were conducted for the analysis and evaluation of the geometry of atomic arrangement. 
However, most studies were based on Monte Carlo, grid-counting, or approximation methods and a high-quality solution requires heavy computational resources, not to mention about its dependency on computation environment. 
In this paper, we introduce a unified framework of computational library, Molecular Geometry Operating System (MGOS), based on an analytic method for the molecular geometry of atomic arrangements.
We believe that the powerful MGOS application programming interface (API) functions will free scientists from developing and implementing complicated geometric algorithms and let them focus on more important scientific problems. 
        \\ \\
{\em Solution method:}
\\
  %Describe the method solution here.
%
Molecular Geometry (MG) is a general framework of mathematical/computational methods for solving molecular structure problems using a geometry-priority philosophy and is implemented by MGOS which is a library of callable C++ API functions.  
MGOS is developed based on the Voronoi diagram of three-dimensional spheres and its two derivative constructs called the quasi-triangulation and beta-complex. 
Note that this Voronoi diagram is different from the ordinary Voronoi diagram of points where the points are atom centers.
Being an analytic method, the solutions of many geometric queries on atomic arrangement, if not all, are obtained correctly and quickly. 
The MGOS architecture is carefully designed in a three-tier architecture so that future modifications and/or improvements can be reflected in the application programs with no additional programming by users.
\\ \\
{\em Additional comments including Restrictions and Unusual features: N/A}\\
  %Provide any additional comments here.
   \\
% \begin{thebibliography}{0}
% \bibitem{1}Reference 1         % This list should only contain those items referenced in the                 
% \bibitem{2}Reference 2         % Program Summary section.   
% \bibitem{3}Reference 3         % Type references in text as [1], [2], etc.
%                               % This list is different from the bibliography at the end of 
%                               % the Long Write-Up.
% \end{thebibliography}
% * Items marked with an asterisk are only required for new versions
% of programs previously published in the CPC Program Library.\\
\end{small}

% \clearpage

\section{Introduction}

In physics, chemistry, and materials science, the properties of inorganic molecules result from the arrangement of their atoms~\cite{MorganEtal04,LiEtal99,FurukawaEtal10}.
In biology, the structure of biomolecules determines their function~\cite{Chothia74,Connolly83,HubbardEtal94,DoyleEtal98,YonathEtal87,LiuTheil05}.
A molecule's properties and interactions with its environment depend
on the geometrical arrangement of its atoms, and geometry has long been one of key issues in the study of atomic arrangements. 
In physics and materials science, examples include the diffusion of lithium
ions through paths closely correlated with geometric
channels~\cite{MorganEtal04}; 
the porosity and surface area of metal organic framework (MOF) for hydrogen storage
\cite{LiEtal99,FurukawaEtal10},
water content regulation in polymer membranes through nanocracks which work as nanoscale valves \cite{ParkEtal16}, to name a few.
In biology, classic examples are the shape complementarity of
the double-helix structure of DNA \cite{GerlingWagenbauerEtal15,OBrienJonesEtal15},
and the
lock-and-key~\cite{Fischer1890} and induced-fit
theories~\cite{Koshland58} of small-molecule binding to
proteins. 
There are many other examples: the linear
relationship between hydrophobic energy and the loss of solvent
accessible surface area~\cite{Chothia74}; the effect of voids on the
solvation and hydration of proteins \cite{HubbardEtal94}; the channel
structure of ion channels and pumps across cell membranes
\cite{DoyleEtal98} and in the ribosome for protein synthesis
\cite{YonathEtal87}; ferritin as a protein nanocage for iron storage
\cite{LiuTheil05}; the Connolly surface of proteins \cite{Connolly83}.
The examples
assert that accurate and efficient geometric computation is critical for understanding and
designing molecules.

However, many studies to date on molecular geometry problems have mostly
been based on Monte Carlo simulation, counting grid points, or
approximations.
For instance, molecular volume is commonly
estimated by counting the numbers of random points or grid points
contained in the molecule \cite{ShrakeRupley73}; conversely, molecular voids are
recognized by removing these grid points \cite{KleywegtJones94}.
Another
example is the imprecise estimation of solvent accessible surfaces
\cite{Richards77}, which is critical
for solvation models used in the calculation of electrostatic energy.

Fig.~\ref{fig:BetaCavityWeb_vs_VOIDOO_benchmark_number_volume_time} shows the comparison between an analytic \cite{KjkEtal14b,KjkEtal15b} 
and a grid-counting \cite{KleywegtJones94} method for computing molecular voids using a test data set consisting of 300 biomolecular structures from the Protein Data Bank (PDB \cite{PDBHome}).
See Appendix I for the 300 PDB codes. 
In Fig.~\ref{fig:BetaCavityWeb_vs_VOIDOO_benchmark_number_volume_time}(a), the horizontal axis denotes the size (i.e. the number of atoms) of each molecule of the test set and
the vertical axis denotes the number of computed voids in the molecular boundary in which at least one water molecule can be placed.
Water molecules are modeled as spherical probes of radius 1.4\AA.
The red filled circle corresponds to the output from the BetaVoid program \cite{KjkEtal14b} 
which implements an analytic method using the Voronoi diagram of three-dimensional spherical atoms.
The other three types of mark denote the results computed by the VOIDOO program (http://xray.bmc.uu.se/usf/voidoo.html) \cite{KleywegtJones94} corresponding to the grid resolutions of 0.1, 0.5, and 1.0.
Fig.~\ref{fig:BetaCavityWeb_vs_VOIDOO_benchmark_number_volume_time}(b) is a zoom-in of the red rectangular box of Fig.~\ref{fig:BetaCavityWeb_vs_VOIDOO_benchmark_number_volume_time}(a).
Note that VOIDOO finds fewer voids than BetaVoid does. 
Fig.~\ref{fig:BetaCavityWeb_vs_VOIDOO_benchmark_number_volume_time}(c) and (d) show the total volume of all the computed voids and
Fig.~\ref{fig:BetaCavityWeb_vs_VOIDOO_benchmark_number_volume_time}(e) and (f) show the computation time taken by the programs.
The following observations were made.
Compared to the correct solutions computed by the BetaVoid program,
VOIDOO finds fewer voids (i.e., it misses many small voids) but significantly overestimates void volumes (despite missing many voids) while it takes significantly more computation time than BetaVoid.
VOIDOO, at 0.1\AA\ grid-resolution, crashes on many moderately sized molecules due to memory shortage.
This experiment clearly shows how an analytic approach compares with an inaccurate and inefficient approach using grid points. 
The experiment was performed on a personal computer with Inter Core i5-4670 CPU (3.4GHz), 8 GB RAM, and Windows 7 Enterprise K (64 bit).

The use of such
resolution-dependent approaches is common despite
their unreliable, inconsistent, and sometimes
conflicting results \cite{VlassiEtal99}.
We observe that VOIDOO is still popular in diverse disciplines \cite{HoriuchiTanakaEtal19,ShimadaKuboEtal17,DucassouDhersEtal17,MolcanSwigonskaEtal17,RizzutoNitschke17,MarkiewiczJenczakEtal17,ChaptalDelolmeEtal17,ZhangBaileyEtal18,KitanovicMarks-FifeEtal18,NagaeYamadaWatanabe18,HtanLuoEtal19} and 
studies of grid-based algorithms continues \cite{ChakravortyGallicchio19}.
The absence of an overarching
analytical theory is because individual researchers have focused on
problem-specific, local aspects of geometry problems, concentrating on
isolated issues such as surfaces,
voids, channels, volumes, areas, and so on.
With so many independently
developed methods, it has been hard to
build a general computational framework
for accurately and efficiently solving all these types of geometrical problems.

\begin{sloppypar}
Here we introduce ``Molecular Geometry (MG)'' as a general framework of mathematical/computational methods
for solving molecular structure problems in geometry-priority approaches, and describe the ``MG Operating System (MGOS)'' which is a library of callable C++ routines for implementing the MG approach in analytical methods.
The proposed analytical methods are based on the Voronoi diagram of three-dimensional spheres \cite{KdsEtal05}, the quasi-triangulation \cite{KdsEtal06,KdsEtal10}, and the beta-complex \cite{KdsEtal10d}.
The MG/MGOS method has three primary advantages: 
application independence,
researcher productivity, and 
solution correctness/accuracy.
In other words, equipped with MG/MGOS, researchers from diverse disciplines can conveniently and easily build computational models to solve molecular geometry problems and quickly obtain correct (or accurate) solutions. 
\end{sloppypar}

\FigSix{BetaCavityWeb_vs_VOIDOO_benchmark_number_volume_time}
{0.47}{0.47}{0.47}{0.47}{0.47}{0.47}
%{0.3}{0.3}{0.3}{0.3}{0.3}{0.3}
{
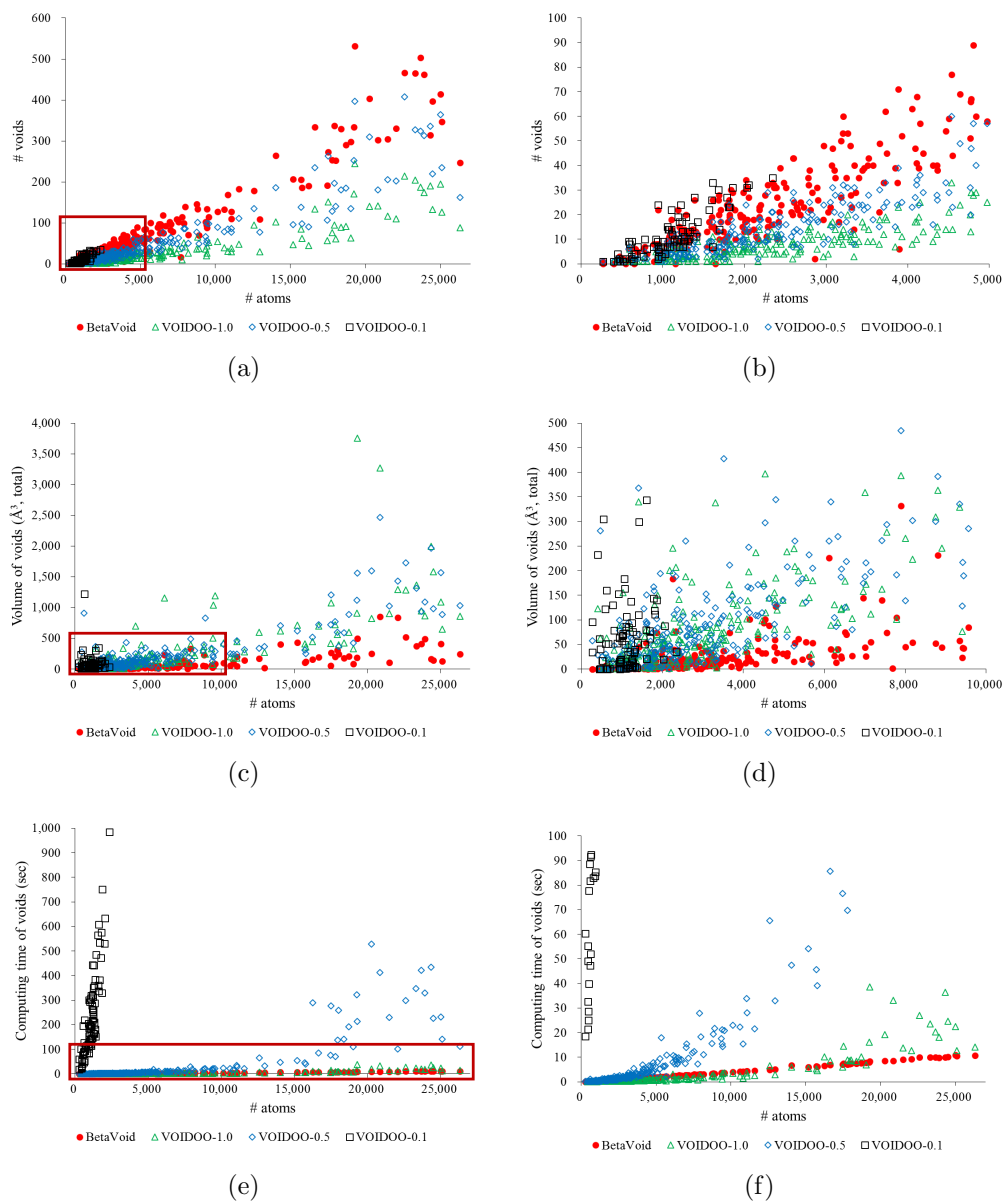
Lee-Richards voids corresponding to a water molecule probe (i.e. 1.4{\AA} radius) computed by BetaVoid \cite{KjkEtal14b} and VOIDOO \cite{KleywegtJones94}. 
The test set consists of 300 PDB structures (Appendix 1 lists the PDB codes).
The red circles correspond to BetaVoid results;
VOIDOO-1.0, VOIDOO-0.5, and VOIDOO-0.1 corresponds to grid resolutions of 1.0, 0.5, and 0.1 {\AA} in VOIDOO, respectively.
The right column is a zoom-in of the left column.
(a) and (b) The number of recognized voids;
(c) and (d) The total volume of the recognized voids;
(e) and (f) Computation time.
}

Section \ref{sec:evolution-of-MG-concept} briefly reviews the evolution of the geometry concepts applied to atomic arrangements for materials and biomolecules.
%XXXXX
Section \ref{sec:overview-of-molecular-geometry} introduces Molecular Geometry as a new computational discipline for studying atomic arrangements.
Section \ref{sec:MGOS-brief} introduces the Molecular Geometry Operating System as a tool for implementing MG. 
Section \ref{sec:use-cases} presents two example molecular geometry problems solved by MGOS. 
Section \ref{sec:MGOS-architecture} presents the application-neutral architecture of MGOS.
Section  \ref{sec:conclusion} concludes. 

\section{How the geometry concept has evolved in the molecular world}
\label{sec:evolution-of-MG-concept}

Johannes Kepler's treatise \emph{The Six-cornered Snowflake} in 1611 and Robert Hooke's book \emph{Micrographia} in 1665 might be the earliest observations of crystallization as a sphere packing process.
In \emph{Cristallographie} in 1783, Rome de L'Isle treated geometry and chemical composition with an equal importance to characterize mineral properties and found ``the law of the constancy of interfacial angles'' which became the foundation of crystallography.
Before the advent of X-ray crystallography, crystals were primarily studied from a geometry perspective.
In 1805, John Dalton introduced the concept of the spherical atom as the indivisible unit of matter and 
in 1874, Le Bel and Van't Hoff independently introduced the concept of tetrahedrally coordinated carbon atoms~\cite{Lebel1874,Vanthoff1874}.
This became the foundation of modern stereochemistry which is the basis of the study of molecular structures~\cite{Grossman89}.
Understanding steric effects (i.e. each atom occupies a certain amount of space) is the basis of the stereochemistry of atoms and provides a geometric understanding of the molecular world.
The coordination number of an atom, defined by Werner in 1893, is still a commonly used geometric measure of atomic arrangement.

In 1940, Sidgwick and Powell proposed that molecular structure is determined by the electron pairs in the valence shell~\cite{SidgwickPowell40,Gillespie72}.
This idea was developed in 1957 by Gillespie and Nyholm ~\cite{GillespieNyholm57} into what is now known as the valence shell electron pair repulsion (VSEPR) model, the name proposed in 1963~\cite{Gillespie63}, which has been used for predicting molecular structure using the Pauli Exclusion Principle, but without solving any explicit equation.
VSEPR is one of the simplest and most successful models of molecular structure ~\cite{Gillespie60,Gillespie63}, and remains popular.
VSEPR can be viewed as a geometric approach to understanding the molecular world.

Molecular biology is the molecular world where geometry has arguably received the most attention.
In 1890, Emil Fischer proposed the well-known lock-and-key theory to explain the interactions between biomolecules.
This is an excellent example of modeling biomolecular phenomena through geometry~\cite{Fischer1890,Fischer1902}.
In 1953, the year that the double-helix structure of DNA was discovered, Francis Crick suggested the idea of a computational approach to the binding between two small molecules through their surfaces~\cite{Crick53a}.
Crick posited that shape complementarity in the helical coiled coil could be modeled as knobs fitting into holes.
This could be the first proposal of explicitly using geometry to understand molecular phenomena, and became the basis of molecular docking.
In 1958, Koshland extended the lock-and-key theory to propose the induced-fit theory~\cite{Koshland58,Koshland63,Koshland94}.

The first determination of the three-dimensional structure of a protein was performed by John Kendrew and Perutz in 1960~\cite{KendrewDickersonEtal60} when they solved the structure of myoglobin. Since then, protein structure determination has become almost routine work; and the PDB contains 152,500 biomolecular structures as of June 8, 2019 \cite{PDBHome}.
Given atomic arrangement databases, such as the PDB, geometry analysis becomes one of the most important research topics for researchers.
Cavities in biomolecules are fundamental for function, stability, dynamics, ligand binding, etc.
The first computational study of cavities in proteins was reported by Lee and Richards in 1971~\cite{LeeRichards71}.
Chothia in 1974 found that the hydrophobic energies in proteins are directly related to the solvent accessible surface area of both polar and non-polar groups, and reported the linear relationship between the hydrophobic energy of proteins and the loss of solvent accessible surface area during folding~\cite{Chothia74,Chothia75}.
This demonstrates that the atoms in folded globular proteins tend to be tightly packed. Thus a large residue volume, and consequently a low overall density, suggests the model of the protein is a poor one and, conversely, a small volume, and high density, suggests is it more likely to be a good one~\cite{Chothia75}.
A protein's interior is closely packed, with few cavities, so that no water molecules are trapped in non-polar cavities~\cite{Chothia75,Richards74}.
The dense packing is critical in stable folding, and residue volumes are directly related to packing energies and conformational entropies.
The stable aggregation of secondary structures increases their interaction area to achieve a high hydrophobicity and results in an increased molecular density.

In the case of enzymes, which are globular proteins, the optimal way of minimizing the volume and the solvent accessible surface area while keeping a constant potential energy is to make the shape as spherical as possible with as few cavities as possible.
Due to the potential energy constraint, the overlap between atoms is limited at a certain level.
Therefore, this is a geometric optimization problem of packing spherical atoms in a spherical container of an appropriate size.
However, certain geometric features need to be conserved for the molecule to maintain its function.
For example, proteasomes require their channel structures for disassembling proteins,
ribosomes need to conserve their channels for synthesizing proteins, while
membrane proteins require channels for the passage of ions.
Therefore, to minimize both volume and accessible surface area under the potential energy constraint, while preserving their crucial geometric features, the interior voids of these proteins must be somehow minimized. 
Hence, the accurate computation of voids in a molecular structure is important for the assessing the structure.
In this regard, the recognition of molecular cavities, such as channels and voids, the computation of their global properties, and understanding their topological structures are fundamental.
As the data in the PDB has been more frequently used, the importance of taking into account its quality has also increased, and there are now a number of tools for assessing structural quality \cite{LaskowskiEtal93,ChenEtal10b,DavisEtal07}. 

\section{Molecular Geometry: A New Approach to Study Atomic Arrangement}
\label{sec:overview-of-molecular-geometry}

%MG PARADIGM
\begin{sloppypar}
Fig. 2 shows the computational process of solving molecular problems. 
In Fig. 2(A), 
Mapping I depicts the
traditional approach of going directly from a particular molecular problem $\mathcal{M}$ to its
solution $Sol(\mathcal{M})$.
There are uncountably many molecular problems and
each problem can have alternative mappings because its modeling is dependent on the nature of the study.
This leads to uncountably many instances of Mapping I.
Each mapping instance usually
consists of nontrivial computational steps and almost always contains a geometry
subproblem involving spherical objects, which in many cases are van der Waals atoms.
Earlier studies \cite{MorganEtal04,LiEtal99,FurukawaEtal10,Chothia74,Connolly83,HubbardEtal94,DoyleEtal98,YonathEtal87,LiuTheil05} show this issue is real and highly common.
Surprisingly, many seemingly easy geometry problems among
spheres remain challenging, if not computationally hard to solve, because of a lack of a suitable
mathematical/computational framework.
Therefore, researchers often spend a significant amount of time and
effort, in the course of solving their geometry problems, developing and implementing their own algorithms. Furthermore,
due to the complexity of
the geometry problems, researchers usually employ Monte Carlo simulation, grid
counting, or other approximate methods.
\end{sloppypar}

\FigOne{molecular-geometry-concept}{1.0}
{
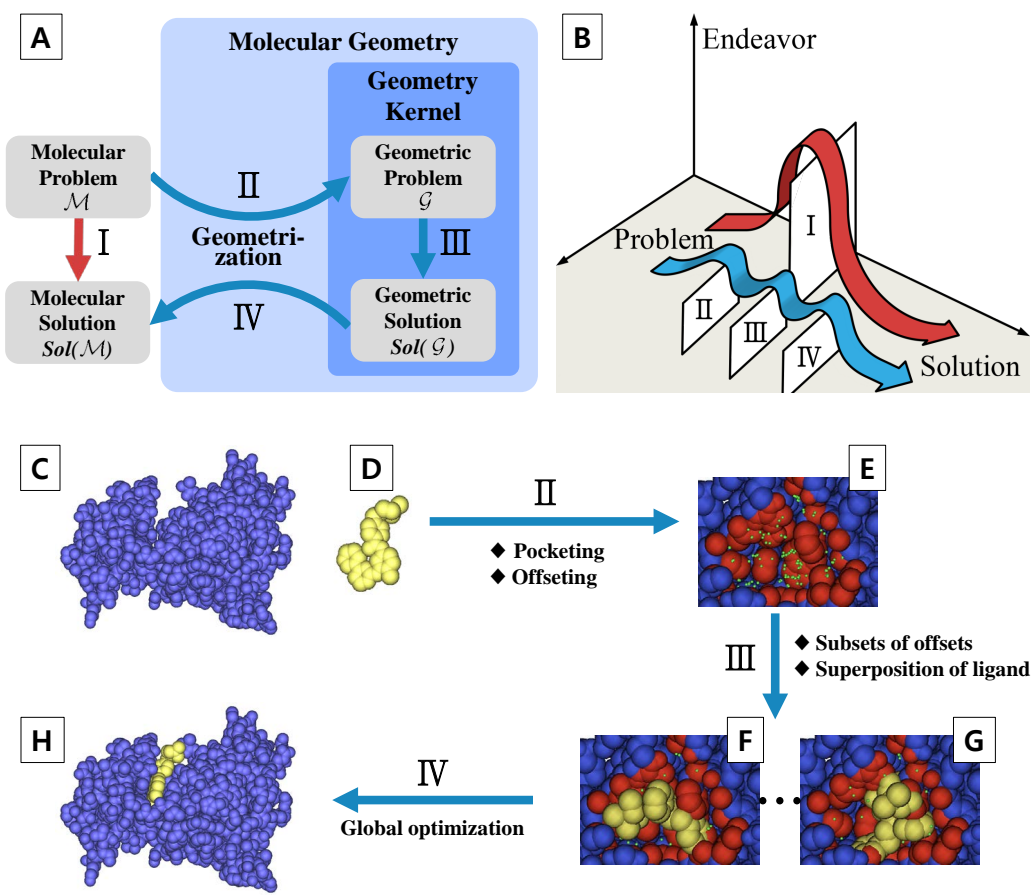
The Molecular Geometry (MG)
framework.  
(A) The MG approach (Mappings II, III and IV) vs. the traditional approach (Mapping I).
(B) The human effort (for developing and implementing algorithms) and computational cost of the MG and traditional approaches.
(C) Receptor.
(D) Ligand.
(E) Pocket.
(F) and (G) Two initial docking poses.
(H) Optimal docking solution.
}

MG provides an alternative, orthogonal method to this traditional approach. It
bypasses the time-consuming and error-prone Mapping I by
taking the walk-around path consisting of Mappings II, III, and IV.
First, the problem $\mathcal{M}$ is modeled as a geometry
problem $\mathcal{G}$ involving spherical atoms (Mapping II). Then, $\mathcal{G}$
is solved via geometric theorems to give the solution
$Sol(\mathcal{G})$ (Mapping III) which is back-transformed to
$\widehat{Sol}(\mathcal{M})$
in
the original molecular space (Mapping IV).
The thesis is that $\widehat{Sol}(\mathcal{M}) \approx Sol(\mathcal{M})$, possibly with
some preconditions. 
The forward and backward transformations of Mappings II and IV are
together called the \emph{geometrization} while the computational methods
for Mapping III form the \emph{geometry kernel}. The
geometrization and geometry kernel together form the basis of the discipline MG
(which is different from the earlier notion \cite{Gillespie72}).

%MG ITERATION

$\widehat{Sol}(\mathcal{M})$
is either close enough to, or a good
approximation of, $Sol(\mathcal{M})$ to allow a more intensive
computational process such as a molecular dynamics (MD) simulation
to be launched. 
As the computational cost of the walk-around path of Mappings II, III, and IV
is significantly cheaper than that of Mapping I, the path may iterate
as many times as necessary by refining the geometrization.  If
the criteria for the convergence of
$\widehat{Sol}(\mathcal{M})$ can be defined, the solution process can
iterate,
possibly without human intervention.  Physicochemical and biological
properties should be carefully reflected during the geometrization.
Given a proper geometrization and a geometry kernel, the path might be automated to iterate if necessary.
Fig. 2(B) 
depicts the significant reduction of both human effort and computational requirement by the MG approach.

Fig. 2(C) 
through (H) illustrate how
a docking simulation program can adopt MG/MGOS in its algorithm.
Given a receptor (C) and a ligand (D) for docking, it is desirable to identify a pocket (E) on the receptor surface where the ligand might bind (Mapping II).
Then, the conformation of the ligand within the pocket can be found by minimizing the distance between the atom sets of both the ligand and the pocket, where the distance is defined by a geometric measure that can be easily evaluated (F and G) (Mapping III) \cite{ShinEtal13}.
Multiple conformations can be found quickly. 
The ligand conformations can then be used as initial solutions for a global optimization procedure such as the genetic algorithm using a  fitness function reflecting the physicochemical and biological measures (H) (Mapping IV). 
It turns out that the geometrical best-fit
solutions using van der Waals radii for atoms are often sufficiently
close to the global solution.
\cite{RjhEtal16} is another example for side-chain prediction. 

%CONDITIONS FOR SUCCESSFUL MG

\begin{sloppypar}
The MG approach has two preconditions: a
mathematically and computationally well-established
geometry kernel and a physicochemically and biologically well-defined geometrization.
The MGOS engine's geometry kernel is written in standard C++ and is based on the Voronoi diagram of three-dimensional spheres \cite{KdsEtal05} and its two derivative constructs \cite{KdsEtal10d}.
The geometrization is inevitably
domain-dependent and is somewhat empirical.
For example, 
different sets of atomic radii may be used for different problems \cite{Bondi64,Slater64}.
%,Pauling47};
The effective Born radius \cite{StillEtal90,OnufrievEtal04} may be most appropriate when using the generalized Born approximation of the Poisson-Boltzmann equation to account for the electrostatic contribution to solvation energy. In studying
a potassium channel's recognition selectivity, its dependence is likely to be on ion radius rather than charge density \cite{LocklessEtal07}. The
%(Structural and Thermodynamic Properties of Selective Ion Binding in a Channel, Steve W. Lockless, Ming Zhou, Roderick MacKinnon);
analysis of protein packing, protein recognition and ligand design \cite{TsaiEtal99}, etc will be governed by the radii of different atomic groups.
Previous studies \cite{MorganEtal04,LiEtal99,FurukawaEtal10,Chothia74,Connolly83,HubbardEtal94,DoyleEtal98,YonathEtal87,LiuTheil05} can be interpreted as efforts at applying different types of geometrization.
A set of geometrization primitives and parameters for each and every application domain should be defined through theoretical studies, experiments, and collaborative thoughts.
\end{sloppypar}

\section{MGOS: The Engine to Implement MG}
\label{sec:MGOS-brief}

MGOS implements MG.
The usefulness of MGOS is akin to a math library for
general-purpose programming languages in science and engineering.
Imagine the time and effort it would take a researcher, even with good programming skills, to code from scratch an algorithm for evaluating, say, $\sin (1.23)$ or $\sqrt{2}$, without a math library.
Would the code be accurate and efficient enough?
Any complicated scientific problem is likely to require calls to many such functions, so 
could one effectively develop an effective program without such a math library?

\begin{sloppypar}
MGOS consists of a set of
natural-language-like application programming interface (API) functions,
easily callable from application programs (see Appendix II for the list of current MGOS APIs) and efficiently provides a correct/accurate solution of geometric queries involving the arrangements of spherical objects where the objects are frequently van der Waals atoms.
For example, the
\texttt{compute\_volume\_and\_area\_of\_van\_der\_Waals\_model()} command computes the volume of the space taken by the atoms (with the van der Waals radii) of a given molecule.
The name of the command is clear about its function.
The \texttt{compute\_voids\_of\_Lee\_Richards\_model()} command
finds all interior voids where an \emph{a priori} defined spherical probe can be placed (e.g. a sphere with 1.4\AA\ radius for water) and computes void properties.
Computed voids can be further processed.
For instance, the voids can be sorted according to volume or boundary area;
the atoms whose boundary contribute to each void can be reported;
the segment of the atom boundary contributing to the void can be
identified and its area computed, etc.
\end{sloppypar}

%COMPUTATIONAL GEOMETRY APPROACH
%
An early attempt at a formal theory to investigate the geometry of atomic arrangement was based on the ordinary Voronoi diagram of points, originally used by Bernal and Finney in 1967 for analyzing liquid structure composed of monosized atoms \cite{BernalFinney67}.
Being the most
compact representation of proximity among points, the ordinary Voronoi
diagram, and its dual called the Delaunay triangulation, has proved the
best method for solving spatial problems for points
\cite{OkabeEtal99}.
%By removing some edge and/or face simplices from
%the Delaunay triangulation, an alpha-shape is obtained to define the
%boundary of the input points with respect to a spherical probe
%\cite{EdelsbrunnerMucke94}.
To extend the theory from points to polysized spheres, we use the
Voronoi diagram of spheres \cite{KdsEtal05}, also called the additively-weighted
Voronoi diagram, which correctly recognizes the Euclidean proximity  among the spherical objects
between any pair of nearby spheres.
Our Voronoi diagram of spheres, along
with its derivative structures, the quasi-triangulation \cite{KdsEtal06,KdsEtal10} and the beta-complex \cite{KdsEtal10d},
provides a powerful computational platform for mathematically rigorous,
algorithmically correct, computationally efficient, and
physicochemically and biologically significant, and practically convenient
method for any geometry problem involving spherical atoms.

\section{Use Cases}\label{sec:use-cases}

We show here how a few simple MGOS APIs can be used to easily compute otherwise difficult to compute geometric features such as voids, channels, water-exposed atoms, etc. of a protein consisting of many atoms.

\subsection{Case I: Analysis of an atomic arrangement}

%Fig.~\ref{fig:void-example} 
Fig. 3 shows a protein structure (PDB id: ijd0) with more than 4,000 atoms. We want to find the boundary atoms exposed to water molecules (modeled as spheres of 1.4\AA~radius), and buried atoms. 
Then, we want to find voids which can contain water molecules and any channel structures that allows the passage of water molecule. 

Fig. 3(A) shows the space-filling, or CPK-model, of the protein structure. 
Observe that there is a tiny hole corresponding to a channel penetrating the structure. 
Fig. 3.(B) shows the quasi-triangulation computed by the MGOS API commands in block B1. 
The command \texttt{MG.preprocess()} computes the Voronoi diagram of the input atoms and transforms it to the quasi-triangulation. 
Fig. 3(C) shows the beta-complex corresponding to water molecules (i.e. spherical probes with 1.4\AA~radius).
Fig. 3(D) and (D') show the atoms exposed to and buried from bulk water, respectively (computed by block B2). 
Hence, the union of the structures in Fig. 3(D) and (D') is the input structure in Fig. 3(A). 
Note that the challenging task of the correct and efficient computation of these structures can be easily and conveniently done by calling a few MGOS APIs. 
Fig. 3(E) shows the voids (green) that may host one or more water molecule (from a geometric point of view) where the molecule is displayed by a ball-and-stick model. 
The voids were computed by the program segment in B3. 
Fig. 3(F) shows the largest (by volume) of the recognized voids, and the atoms whose boundaries contribute to the boundary of this void. 
We call these atoms the contributing atoms.
%The biggest void is the one with the largest volume. 
%
If it is necessary to investigate if a water molecule can indeed be placed in the void, the biochemical or biophysical properties of the surface segments of the void boundary can be further analyzed by computing the precise geometric information of the patches of atomic boundaries using MGOS APIs. 
%
%MGOS has such APIs. 
%
In fact, the \texttt{compute\_voids\_of\_Lee\_Richards\_model( WATER\_SIZE )} finds all voids that may contain water molecule(s), computes the volume of each void, computes the boundary area of each void, finds the contributing atoms, computes the area of the contributing patch(es) of each contributing atom, etc. 
The program segment in B4 simply returns the contributing atom information already computed by the command above. 
Fig. 3(G) shows the channels that may allows a water molecule to move. 
Like the voids, the surface properties of these channels can be further investigated if necessary. 
These channels were computed by the program segment in B5. 
Fig. 3(H) and (I) show two different visualizations of the biggest channel with its contributing atoms and spine, respectively. 
This biggest channel is located by the program segment in B6.
Refer to Supplementary Video 1 for the three-dimensional animation of this computational process.

\FigOne{void-example}{0.75}
{
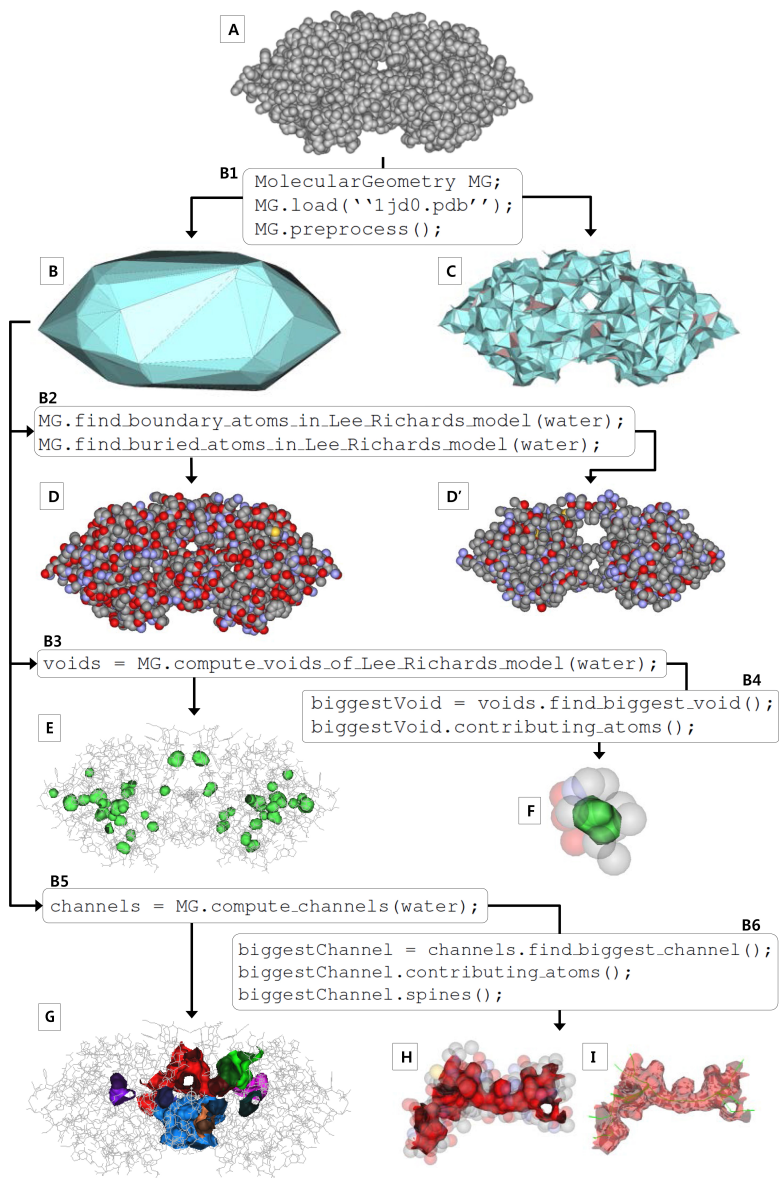
Protein structure analysis using \texttt{Program-Use-Case-I} in Fig.~\ref{fig:sample-program-1} (PDB code: 1jd0; 4,195 atoms).
See Supplementary Video 1.
(A) and (B) The space-fillin and quasi-triangulation models of the input structure.
(C) The beta-complex for water (i.e. a spherical probe with radius 1.4\AA).
(D) and (D') The atoms exposed to and buried from bulk water,  respectively. 
(E) The (green) voids for water.
(F) The largest void and its contributing atoms. 
(G) The channels for water. 
(H) and (I) Two different visualizations of the biggest channel with its contributing atoms and spine, respectively. }

\begin{figure*}[hp!t!b] % here present top bottom
\begin{center}
\begin{boxedminipage}{0.99\textwidth}
\scriptsize

\begin{tabbing}
\= \kill
\> /* \texttt{Program-Use-Case-I}: Analysis of single atomic arrangement */ \\
000 \= \kill
%   \> // \texttt{Program-Use-Case-I} \\
   \> // include a head file for using MGOS\\
1  \> \texttt{\#include "MolecularGeometry.h" }\\
2  \> \texttt{using namespace MGOS; }\\
   \> // define the probe size for water molecule\\
3  \> \texttt{const double WATER\_SIZE = 1.4;}\\
\\
4  \> \texttt{int main() }\\
5  \> \texttt{\{ }\\
000 \= 000 \= 00000 \= \kill
   \> \> // B1: load a PDB model and preprocess\\
6  \> \> \texttt{MolecularGeometry MG; }\\
7  \> \> \texttt{MG.load( "1jd0.pdb" ); }\\
8  \> \> \texttt{MG.preprocess(); }\\
\\
   \> \> // B2: find the atoms on boundary and of interior corresponding to water molecule\\
9  \> \> \texttt{AtomPtrSet boundaryAtoms = MG.find\_boundary\_atoms\_in\_Lee\_Richards\_model( WATER\_SIZE ); }\\
10 \> \> \texttt{AtomPtrSet interiorAtoms = MG.find\_buried\_atoms\_in\_Lee\_Richards\_model( WATER\_SIZE ); }\\
\\
   \> \> // B3: compute the voids inside the protein which are defined by water molecule size\\
11 \> \> \texttt{MolecularVoidSet voids = MG.compute\_voids\_of\_Lee\_Richards\_model( WATER\_SIZE ); }\\
\\
   \> \> // B4: find the biggest void and its contributing atoms\\
12 \> \> \texttt{MolecularVoid biggestVoid = voids.find\_biggest\_void();}\\
13 \> \> \texttt{AtomPtrSet contributingAtomsOfBiggestVoid = biggestVoid.contributing\_atoms();}\\
\\
   \> \> // B5: compute the channels inside the protein which are defined by water molecule size\\
14 \> \> \texttt{MolecularChannelSet channels = MG.compute\_channels( WATER\_SIZE ); }\\
\\
   \> \> // B6: find the biggest channel, its contributing atoms, and spines\\
15 \> \> \texttt{MolecularChannel biggestChannel = channels.find\_biggest\_channel();}\\
16 \> \> \texttt{AtomPtrSet contributingAtomsOfBiggestChannel = biggestChannel.contributing\_atoms();}\\
17 \> \> \texttt{biggestChannel.spine();}\\
\\
18 \> \> \texttt{return 0;}\\
19 \> \texttt{\}}
\end{tabbing}

\end{boxedminipage}
\end{center}
\caption{\texttt{Program-Use-Case-I} computes the voids, channels, etc. in Fig. 3}
\label{fig:sample-program-1}
\end{figure*}

\texttt{Program-Use-Case-I} in Fig.~\ref{fig:sample-program-1} is 
the complete code of an application program which embeds the MGOS APIs to perform the required computation.
%Fig.~\ref{fig:void-example}.
%
The first line of the code includes the \texttt{MolecularGeometry.h}
file which defines the MGOS classes to be used by the program.  Line 7
loads an input file of PDB format. The \texttt{MG.preprocess()}
command in line 8 computes the Voronoi diagram of the input structure
and transforms it to its quasi-triangulation.
If the quasi-triangulation file already exists in the working directory, this command directly loads
the file.

\begin{sloppypar}
The command
\texttt{MG.find\_boundary\_atoms\_in\_Lee\_Richards\_model()} in line
9 finds the set of boundary atoms of the Lee-Richards solvent
accessible model where the solvent is represented as a spherical probe
for water with the radius 1.4\AA , as defined in line 3.
Similarly, \texttt{MG.find\_buried\_atoms\_in\_Lee\_Richards\_model()}
finds the set of buried atoms of the Lee-Richards solvent accessible
model.  It is worth noting that without the MGOS engine, it is very
difficult to correctly and efficiently find these sets because it is
necessary to distinguish the atoms exposed to solvent from those that are
buried.
\end{sloppypar}

\begin{sloppypar}
The command \texttt{MG.compute\_voids\_of\_Lee\_Richards\_model()} in
line 11 locates all voids inside the Lee-Richards solvent accessible
model.  After finding the voids, this command computes the geometric
properties such as the volume and area of each void.  Then,
\texttt{voids.find\_biggest\_void()} in line 12 finds the void
with the biggest volume.  The command
\texttt{biggestVoid.contributing\_atoms()} in the next line finds all
the atoms contributing to the boundary of the biggest void.
\end{sloppypar}

%\begin{sloppypar}
\texttt{MG.compute\_channels()} in line 14 computes all of the channels inside the Lee-Richards solvent accessible model.
\texttt{channels.find\_biggest\_channel()} in line 15 finds the biggest channel and the atoms contributing to the boundary of this biggest channel are given by\\ 
\texttt{biggestChannel.contributing\_atoms()}.
\texttt{biggestChannel.spine()} in line 17 finds the spine of the biggest channel.
%\end{sloppypar}

Fig. 5, 
together with Supplementary Videos 2, 3, and 4, show other examples of voids and channels that can be recognized by a slight modification of the \texttt{Program-Use-Case-I} code with ferritin, a potassium channel, and a metal-organic framework.

\FigOne{proteasome-ferritin-mof}{1.0} 
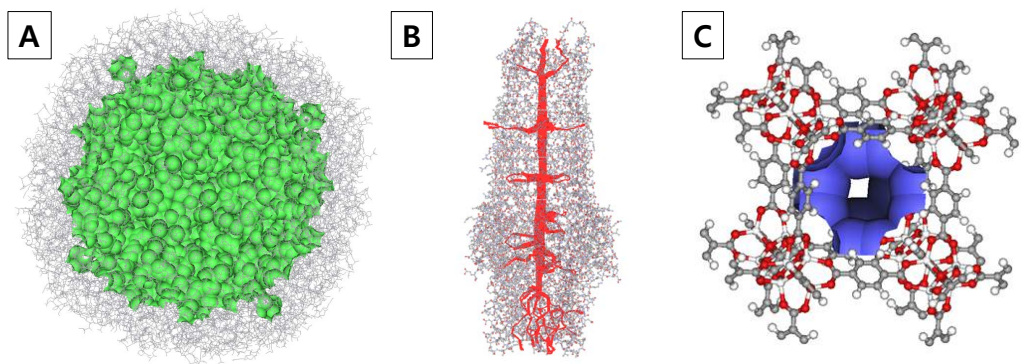
{Geometric features in
  atomic arrangements shown in ball-and-stick diagram.
  See Supplementary Video 2, 3, and 4.
  (A) Ferritin (PDB code: 1mfr; 34,320 atoms). 
  The largest void (green) for water (modeled as a sphere of 1.4\AA\ radius). 
  492 tiny voids are additionally found but are not shown here because they are biologically insignificant.
  (B) Potassium channel protein (PDB code: 2vdd; 9,915 atoms) showing a potassium channel (red).
  (C) Metal organic framework MOF5 \cite{LiEtal99} and its Lee-Richards accessible surface (blue) corresponding to a 2.0\AA\ spherical probe. The geometric properties such as the pore volume, apparent surface area, etc. are critical for MOF design.
  }

\clearpage
\subsection{Case II: Analysis of 100 atomic arrangements}

Fig.~\ref{fig:sample-program-2} shows the code of another application program, \texttt{Program-Use-Case-II}, which
analyzes multiple PDB files to compute the volumes, areas, and voids of 100
molecular structures (arbitrarily selected for demonstration purpose) together with the computation time statistics.

\begin{figure*}[hp!t!b] % here present top bottom
\begin{center}
\begin{boxedminipage}{0.99\textwidth}
\scriptsize

\begin{tabbing}
\= \kill
\> /* \texttt{Program-Use-Case-II}: Analysis of multiple atomic arrangements */ \\
000 \= \kill
   \> // include a head file for using MGOS\\
1  \> \texttt{\#include "MolecularGeometry.h" }\\
2  \> \texttt{using namespace MGOS; }\\
   \> // include a head file for using file I/O functions\\
3  \> \texttt{\#include "MGUtilityFunctions.h"}\\
   \> // define the extension of PDB file.\\
4  \> \texttt{const string PDBFileExtension(".pdb");}\\
5  \> \texttt{int main(int argc, char* argv[]) }\\
6  \> \texttt{\{ }\\
000 \= 000 \= \kill
    \> \> // open a file which contains the list of PDB codes\\
7   \> \> \texttt{ifstream fileFor100PDBCodes( argv[ 1 ] );}\\
    \> \> // get the number of PDB files\\
8   \> \> \texttt{int numberOfPDBFiles = get\_the\_number\_of\_PDB\_files( fileFor100PDBCodes );}\\
    \> \> // get the solvent probe radius\\
9   \> \> \texttt{double solventProbeRadius = atof( argv[ 2 ] );}\\
    \> \> // open a blank file for writing the result\\
10   \> \> \texttt{ofstream fileForMassPropertyAndVoidStatistics( argv[ 3 ] );}\\
    \> \> // write the column names in the first line of the file\\
11  \> \> \texttt{write\_column\_names\_of\_output\_file( fileForMassPropertyAndVoidStatistics );}\\
    \> \> // process each and every PDB model\\
12  \> \> \texttt{int i\_PDB;}\\
13  \> \> \texttt{for (i\_PDB = 0;i\_PDB < numberOfPDBFiles;i\_PDB++)}\\
14  \> \> \texttt{\{}\\
000 \= 000 \= 000 \= \kill
    \> \> \> // get current PDB code\\
15  \> \> \> \texttt{string currentPDBCode = get\_current\_PDB\_code( fileFor100PDBCodes );}\\
\\
    \> \> \> // load a PDB model, preprocess, and measure elapsed time\\
16  \> \> \> \texttt{MolecularGeometry MG;}\\
17  \> \> \> \texttt{MG.load( currentPDBCode + PDBFileExtension );}\\
18  \> \> \> \texttt{MG.preprocess();}\\
19  \> \> \> \texttt{double timeForVDQT = MG.elapsed\_time();}\\
\\
    \> \> \> // compute volume and area, and measure elapsed time\\
20  \> \> \> \texttt{MolecularMassProperty massProperty}\\
21  \> \> \> \texttt{= MG.compute\_volume\_and\_area\_of\_Lee\_Richards\_model( solventProbeRadius );}\\
22  \> \> \> \texttt{double timeForVolumeAndArea = MG.elapsed\_time();}\\
\\
   \> \> \> // compute void, and measure elapsed time\\
23  \> \> \> \texttt{MolecularVoidSet voids}\\
24  \> \> \> \texttt{= MG.compute\_voids\_of\_Lee\_Richards\_model( solventProbeRadius );}\\
25  \> \> \> \texttt{double timeForVoid = MG.elapsed\_time();}\\
\\
   \> \> \> // write the statistics of current model\\
26  \> \> \> \texttt{write\_statistics\_for\_current\_PDB\_model(currentPDBCode, MG, massProperty, voids, timeForVDQT,}\\
27  \> \> \> \texttt{timeForVolumeAndArea, timeForVoid, fileForMassPropertyAndVoidStatistics);}\\
28  \> \> \texttt{\}}\\
29 \> \> \texttt{return 0;}\\
30 \> \texttt{\}}
\end{tabbing}

\end{boxedminipage}
\end{center}
\caption{\texttt{Program-Use-Case-II} that computes the volumes, areas, and voids of 100 molecular structures }
\label{fig:sample-program-2}
\end{figure*}

\clearpage
\texttt{Program-Use-Case-II} requires four pieces of input data:
(i) A file containing PDB codes (Fig. \ref{fig:input-file-out-file-for-sample-program-2-a}),
(ii) the size of the solvent probe,
(iii) the output file to store computed results (Fig. \ref{fig:input-file-out-file-for-sample-program-2-b}), and
(iv) the PDB model files.

\begin{figure*}[hp!t!b] % here present top bottom
\begin{center}
\begin{scriptsize}

\subfigure[]{
\begin{boxedminipage}{0.3\textwidth}
\texttt{100} \\
\texttt{1c26}\\
\texttt{1d2k}\\
$\vdots$ \\
\texttt{4eug}
\end{boxedminipage}
\label{fig:input-file-out-file-for-sample-program-2-a}
}

\subfigure[]{
\begin{boxedminipage}{0.81\textwidth}
\begin{tabular}{@{}c@{}r@{}r@{}r@{}r@{}r@{}r@{}r@{}}
\texttt{ code,}  & \texttt{ \#atoms,} & \texttt{ volume, } & \texttt{ area   ,} & \texttt{ \#voids,} & \texttt{ T(VD/QT),} & \texttt{ T(mass),} & \texttt{ T(void)}\\
\texttt{ 1c26,}  & \texttt{    268, } & \texttt{ 7410.01,} & \texttt{ 2393.33,} & \texttt{       0,} & \texttt{     514 ,} & \texttt{        110,} & \texttt{         78}\\
\texttt{ 1d2k,}  & \texttt{    3082,} & \texttt{ 66165.1,} & \texttt{ 4179.02,} & \texttt{      47,} & \texttt{    7990 ,} & \texttt{        952,} & \texttt{        936}\\
\texttt{      }  & \texttt{          } & \texttt{        } & \texttt{         } & \texttt{         } & \texttt{          } & \texttt{            } & \texttt{           }\\
\texttt{ 4eug,}  & \texttt{    1788,} & \texttt{ 39124.8,} & \texttt{ 3710.06,} & \texttt{      11,} & \texttt{    3748 ,} & \texttt{        484,} & \texttt{        406}
\end{tabular}
\end{boxedminipage}
\label{fig:input-file-out-file-for-sample-program-2-b}
}

\end{scriptsize}

\end{center}

\caption{Examples of files for \texttt{Program-Use-Case-II}: 
(a) The input file storing the 100 PDB codes.
(b) The output file for computation result.}
\label{fig:input-file-out-file-for-sample-program-2}

\end{figure*}

\begin{sloppypar}
The program begins by including \texttt{MGUtilityFunctions.h} in
addition to \texttt{MolecularGeometry.h} because the program also uses some
utility functions related to file I/O.  
The command in line 7
opens a file, say FILE\_{IN}, which contains the 100 PDB codes to use.
The first line of FILE\_{IN} contains the number 100 of PDB models.
Each of the following lines contains a PDB code as shown in Fig.
\ref{fig:input-file-out-file-for-sample-program-2}(a).  The next
command \texttt{get\_the\_number\_of\_PDB\_files()} returns ``100'' by
referring to the first line of FILE\_{IN}.  Line 9 sets the size
of the solvent probe from the command line invoking program execution.
Line 10 opens a blank output file, say FILE\_{OUT}, for the
computed results.  The command
\texttt{write\_column\_names\_of\_output\_file()} writes the column
names to the first line of FILE\_{OUT} as shown in Fig.
\ref{fig:input-file-out-file-for-sample-program-2}(b).
\end{sloppypar}

\begin{sloppypar}
The code chunk in lines 12--28 processes each PDB model by computing
geometric features, measuring the elapsed times, and writing the
results to FILE\_{OUT}.  Line 15 gets the current PDB code from
FILE\_{IN} and the corresponding PDB model is loaded by line 17.
After the model is preprocessed in line 18, the elapsed time is given
by the command \texttt{MG.elapsed\_time()} in the next line.
\texttt{MG.compute\_volume\_and\_area\_of\_Lee\_Richards\_model()} in
line 21 computes the volume and area for the Lee-Richards solvent
accessible model.  The program also counts the elapsed time in the
next line.  Similarly,
\texttt{MG.compute\_voids\_of\_Lee\_Richards\_model()} in line 24
computes the voids for the solvent molecule and then, the elapsed time
is counted.  The command
\texttt{write\_statistics\_for\_current\_PDB\_model()} in lines 26 and
27 writes the statistics for the current PDB model, such as PDB code, the
number of atoms, volume, area, the number of voids, and time
statistics as in Fig.
\ref{fig:input-file-out-file-for-sample-program-2}(b).  The code for
\texttt{MGUtilityFunctions} used in \texttt{Program-Use-Case-II} is shown
in Fig. \ref{fig:MGUtilityFunctions}.
\end{sloppypar}

\begin{figure*}[hp!t!b] % here present top bottom
\begin{center}
\begin{boxedminipage}{0.99\textwidth}
\scriptsize

\begin{tabbing}
\= \kill
\> /* \texttt{MGUtilityFunctions} */ \\
000 \= \kill
1  \> \texttt{\#include "MGUtilityFunctions.h"}\\
%\\
%2  \> \texttt{ifstream open\_file\_for\_list\_of\_PDB\_codes(char* fileName)}\\
%3  \> \texttt{\{ }\\
000 \= 000 \= \kill
%4  \> \> \texttt{ifstream fileFor100PDBCodes( fileName );}\\
%5  \> \> \texttt{return fileFor100PDBCodes;}\\
%6  \> \texttt{\}}\\
\\
2  \> \texttt{int get\_the\_number\_of\_PDB\_files(ifstream\& fileFor100PDBCodes)}\\
3  \> \texttt{\{ }\\
   \> \> // get the number of PDB files by interpreting the first line as an integer \\
4  \> \> \texttt{const int bufferSize = 100;}\\
5  \> \> \texttt{char line[bufferSize];}\\
6 \> \> \texttt{fileFor100PDBCodes.getline(line, bufferSize);}\\
7 \> \>  \texttt{char* delims = "\textbackslash n \textbackslash t";}\\
8  \> \> \texttt{char* token  = strtok(line, delims);}\\
9 \> \> \texttt{int numberOfPDBFiles  = std::atoi( token );}\\
10 \> \> \texttt{return numberOfPDBFiles;}\\
11 \> \texttt{\}}\\
\\
%17 \> \texttt{ofstream open\_file\_for\_writing\_result(char* fileName)}\\
%18 \> \texttt{\{ }\\
%19 \> \> \texttt{ofstream fileForMassPropertyAndVoidStatistics( fileName );}\\
%20 \> \> \texttt{return fileForMassPropertyAndVoidStatistics;}\\
%21 \> \texttt{\}}\\
%\\
12 \> \texttt{void write\_column\_names\_of\_output\_file(ofstream\& fileForMassPropertyAndVoidStatistics)}\\
13 \> \texttt{\{ }\\
14 \> \> \texttt{fileForMassPropertyAndVoidStatistics $\ll$ "PDBFileName,"} \\
15 \> \> \texttt{$\ll$ "numberOfAtoms," $\ll$ "volume," $\ll$ "area," $\ll$ "numberOfVoids,"}\\
16 \> \> \texttt{$\ll$ "time(VD/QT)," $\ll$ "time(mass)," $\ll$ "time(void)" $\ll$ endl;}\\
17 \> \texttt{\}}\\
\\
18 \> \texttt{string get\_current\_PDB\_code(ifstream\& fileFor100PDBCodes)}\\
19 \> \texttt{\{ }\\
   \> \> // get the current PDB code by interpreting the current line as a string \\
20 \> \> \texttt{const int bufferSize = 100;}\\
21 \> \> \texttt{char line[bufferSize];}\\
22 \> \> \texttt{fileFor100PDBCodes.getline(line, bufferSize);}\\
23 \> \> \texttt{char* delims = "\textbackslash n \textbackslash t";}\\
24 \> \> \texttt{char* token  = strtok(line, delims);}\\
25 \> \> \texttt{string currentPDBCode = string( token );}\\
26 \> \> \texttt{return currentPDBCode;}\\
27 \> \texttt{\}}\\
\\
28 \> \texttt{void write\_statistics\_for\_current\_PDB\_model( }\\
29 \> \texttt{const string\& currentPDBCode, MolecularGeometry\& MG,}\\
30 \> \texttt{MolecularMassProperty\& massProperty, MolecularVoidSet\& voids,}\\
31 \> \texttt{const double\& timeForVDQT, const double timeForVolumeAndArea,}\\
32 \> \texttt{const double timeForVoid, ofstream\& fileForMassPropertyAndVoidStatistics)}\\
33 \> \texttt{\{ }\\
   \> \> // get the statistics such as the number of atoms, volume, area, and the number of voids \\
34 \> \> \texttt{int numberOfAtoms = MG.number\_of\_atoms();}\\
35 \> \> \texttt{double volume = massProperty.volume();}\\
36 \> \> \texttt{double area   = massProperty.area();}\\
37 \> \> \texttt{int numberOfVoids = voids.number\_of\_voids();}\\
\\
   \> \> // write the statistics in the output file \\
38 \> \> \texttt{fileForMassPropertyAndVoidStatistics $\ll$ currentPDBCode $\ll$ ","} \\
39 \> \> \texttt{$\ll$ numberOfAtoms $\ll$ "," $\ll$ volume $\ll$ "," $\ll$ area $\ll$ "," $\ll$ numberOfVoids $\ll$ ","}\\
40 \> \> \texttt{$\ll$ timeForVDQT $\ll$ "," $\ll$ timeForVolumeAndArea $\ll$ "," $\ll$ timeForVoid $\ll$ endl;}\\
41 \> \texttt{\}}
\end{tabbing}

\end{boxedminipage}
\end{center}
\caption{\texttt{MGUtilityFunctions} which are used in \texttt{Program-Use-Case-II}.}
\label{fig:MGUtilityFunctions}
\end{figure*}

\begin{sloppypar}
Fig. \ref{fig:analysis-of-produced-data} shows the graphs produced by using Microsoft Excel with the output file FILE\_{OUT} for some computed results for the 100 PDB files.
Fig. \ref{fig:analysis-of-produced-data}(a), (b), and (c) are the volumes, areas, and the numbers of voids, respectively.
Fig. \ref{fig:analysis-of-produced-data}(d) shows the time for computing the Voronoi diagram and quasi-triangulation.
Fig. \ref{fig:analysis-of-produced-data}(e) is the time for computing the volume and area, and
Fig. \ref{fig:analysis-of-produced-data}(f) for the voids.
Note that these graphs can be produced by a few clicks of the mouse button and column choices.
\end{sloppypar}

\FigSix{analysis-of-produced-data}
{0.45}{0.45}{0.45}{0.45}{0.45}{0.45}
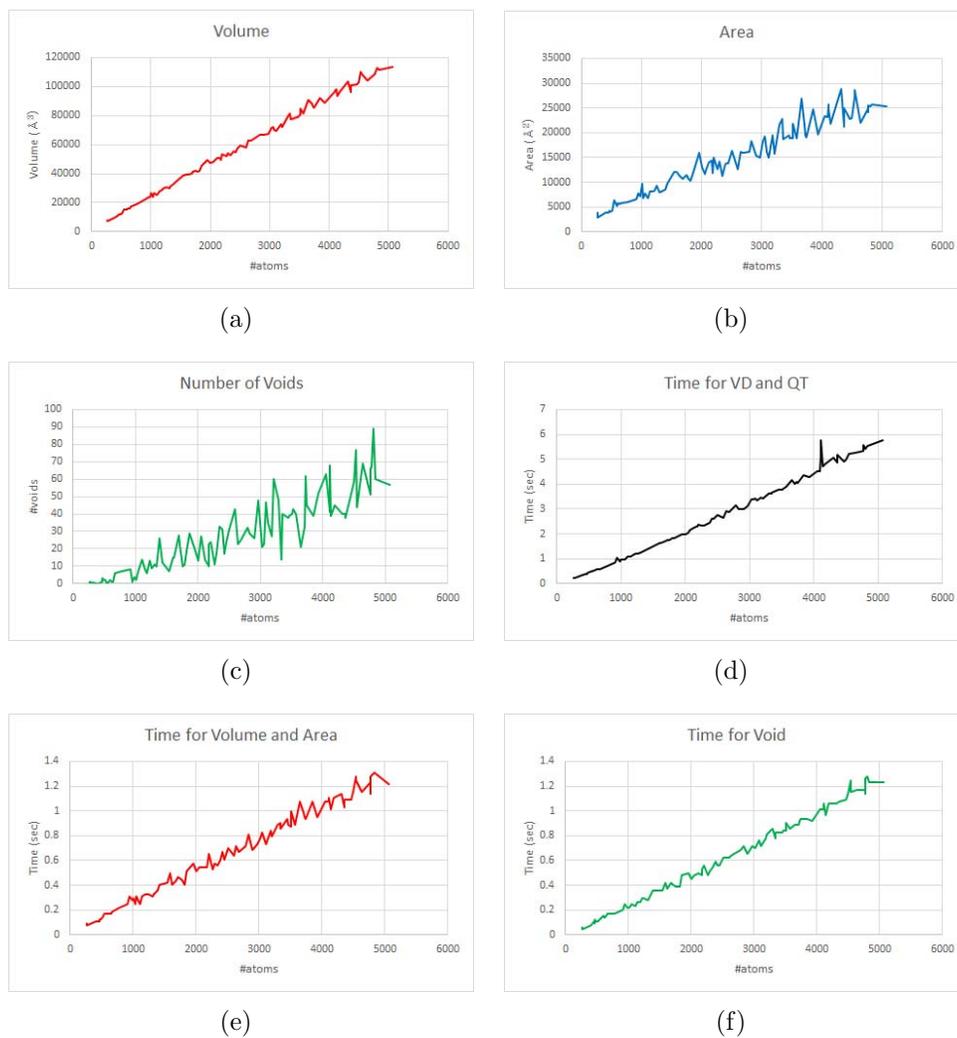
{Graphs produced by Microsoft Excel using the output file FILE\_{OUT}. Volume, area, the number of voids, and time statistics for the 100 PDB models, ordered by the total number of atoms:
(a) volumes,
(b) areas,
(c) the numbers of voids,
(d) time for computing Voronoi diagram and its quasi-triangulation,
(e) time for volumes and areas, and
(f) time for voids.}

\section{MGOS Architecture}\label{sec:MGOS-architecture}

\begin{sloppypar}
The architecture of MGOS has been carefully designed so that any future
modifications will not require rewriting the existing code of an application program.
If a molecular problem can be properly geometrized in terms of
appropriate-sized spherical balls, the MG/MGOS framework can quickly provide the best possible
solution.
It is expected that the MGOS engine will evolve, with new functions to be added in the future.
One area of interest is in developing methods for the optimal design of molecules in the concept of ``operating system'', e.g. in terms of side chain conformations, to develop a program to help engineer proteins.
\end{sloppypar}

{\bf Software architecture:}
MGOS is middleware, connecting application programs with a
low-level Geometry Library performing geometric
computations (Fig. 10).
It is composed of a set of API functions callable from application programs;
each is implemented by calls to the Geometric Library's functions which are
application-independent. In addition to geometric properties, MGOS also makes use of molecular properties such as force-fields, electostatics, etc.

The Geometric Library, the application-independent low level library, performs geometric computations among spherical objects and is based on three closely related constructs: the Voronoi diagram of three-dimensional spheres, the quasi-triangulation, and the beta-complex.\\

\FigOne{architecture-of-MGOS}{0.9}
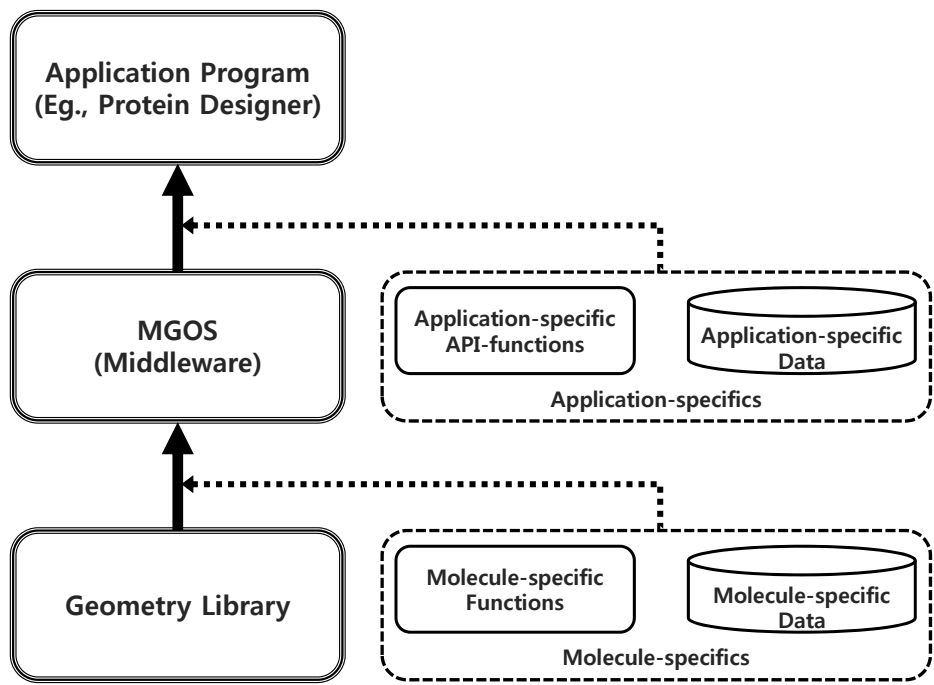
{The role of MGOS for creating application programs.
MGOS is a middleware engine connecting application programs to a set of appropriate API-functions where each performs the required geometric computation.}

{\bf Topology data structure:}
Fig. 11 shows the design of the fundamental data structure for topology in the MGOS library.
Three types of Voronoi diagrams (i.e., the ordinary Voronoi diagram of points, the power diagram, and the Voronoi diagram of spherical balls) are all stored in the radial-edge data structure (REDS) which is appropriate to represent cell-structured non-manifold objects \cite{Lee99}.
``\texttt{REDS}'' in the figure is a member data of the Voronoi diagram itself, which is denoted by \texttt{VoronoiDiagram}.
On the other hand, the dual structure is denoted by \texttt{Triangulation} and has three instances (i.e., the Delaunay triangulation, the regular triangulation, and the quasi-triangulation which are respective dual structures of the three types of Voronoi diagrams above) and is stored in the inter-world data structure (IWDS) \cite{KdsEtal06}.
``\texttt{IWDS}'' in the figure is a member data of the triangulation itself, denoted by \texttt{Triangulation}.

The dual transformation is implemented between the two classes of REDS and IWDS.
Thus the three dual transformations
(i.e., the dual transformation between the ordinary Voronoi diagram of points and the Delaunay triangulation,
that between the power diagram and the regular triangulation, and
that between the Voronoi diagram of spheres and the quasi-triangulation)
are all implemented through the transformation between REDS and IWDS.
All three transformation instances are facilitated by a single transformation as they are all stored in the same topology data structure.

\FigOne{class_design_core-tier}{1}
{
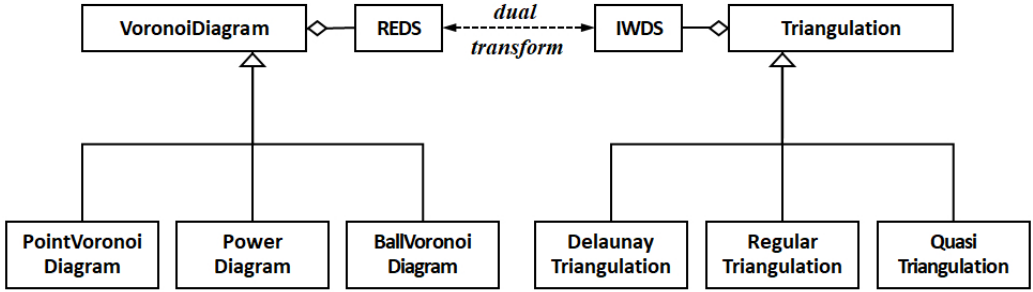
Class design of the topology structures in MGOS.
The dual transformation between REDS and IWDS facilitates those between all three types of Voronoi diagrams and all three types of triangulations.
}

Fig. 12 shows the details of REDS and IWDS. 
REDS in Fig. 12(a) stores the topology of the Voronoi diagram and has the class definitions of the topological entities of the Voronoi diagram:
cells, faces, edges, and vertices which are denoted by $\textrm{VD\_Cell}$, $\textrm{VD\_Face}$, $\textrm{VD\_Edge}$, and $\textrm{VD\_Vertex}$. 
Each cell points to $|F|$ faces which define its boundary and each face points to two incident cells. 
Each vertex points to its four incident edges and each edge points to its two vertices. 
In the ordinary  Voronoi diagram of points, or power diagram, a face has only one loop of edges which defines the boundary of the face (thus called the outer-loop). 
In the Voronoi diagram $\mathcal{VD}$ of 3D spheres, however, a face may have an inner-loop(s) in addition to the outer-loop where each corresponds to an edge-graph disconnected from that of the rest of the entire Voronoi diagram.
This observation is reflected in the pointer from $\textrm{VD\_Face}$ to $\textrm{Loop}$. 
In the Voronoi diagram, an edge has three, and only three, incident faces and in REDS, each edge has three copies of its replica called partial edges $\textrm{PartialEdge}$ where each participates in the loop of an incident face. 
The three partial edges are connected in a circular manner in the counterclockwise orientation around the directed $\textrm{VD\_Edge}$ and in our implementation, each $\textrm{VD\_Edge}$ points one of the partial edges. 

IWDS in Fig. 12(b) stores the topology of the quasi-triangulation and has the class definitions of topological entities of cells, faces, edges, and vertices of the quasi-triangulation
which are denoted by $\textrm{QT\_Cell}$, $\textrm{QT\_Face}$, $\textrm{QT\_Edge}$, and $\textrm{QT\_Vertex}$. 
Note that the data structure is designed for the quasi-triangulation because the other two triangulations are its special cases.
In the quasi-triangulation, a cell has four faces and each face has two incident cells;
A face has three edges and an edge has a set of $1+N_{small-world}$ pointers where each of the $N_{small-world}$ pointers indicate the entrance to a small-world which corresponds to an inner loop of $\textrm{QT\_Face}$.
An edge has two vertices and a cell has four vertices.
A vertex has a pointer to an incident edge and one to an incident cell.

\FigTwo{data_structure_REDS_IWDS}{0.7}{0.7}
{
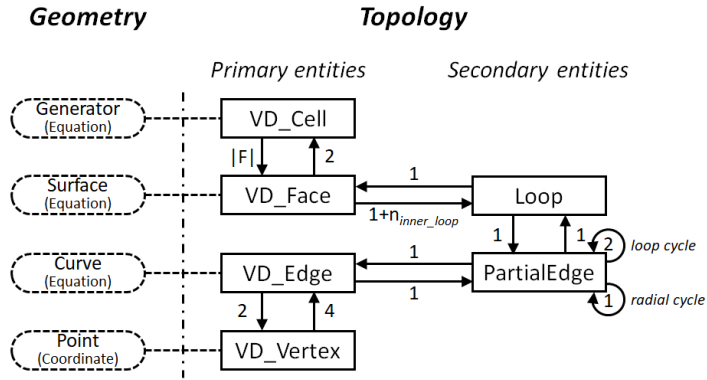
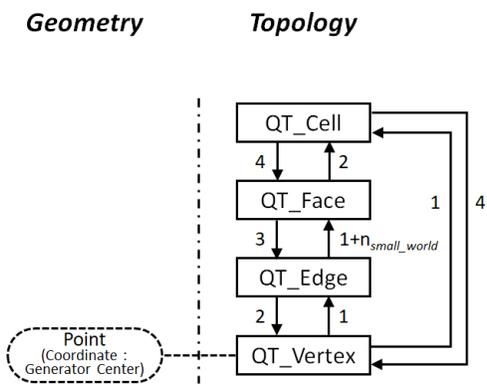
Data structure of REDS and IWDS.
(a) REDS,
(b) IWDS.
}

\begin{comment}
\color{red}
\subsection{Computation efficiency and Walk-around} 

COMPUTATION TIME....

BIG MODEL....

VD --- QT....SYSTEM DESIGN

QTDB....

\color{black}

\end{comment}

\section{Conclusions}\label{sec:conclusion}

Despite the importance of the geometry of atomic arrangements in many fields, no general framework of mathematical/computational theory for the geometry of atomic arrangement exists.
In this paper, we introduce ``Molecular Geometry (MG)'' as a theoretical framework and ``MG Operating System (MGOS)'' as a middleware to implement the MG theory.
\begin{sloppypar}
We assert that MG/MGOS will free researchers from
time-consuming and error-prone tasks of developing and implementing
highly sophisticated and complex algorithms of a geometrical nature for molecular structure studies
so that they can focus more on
fundamental research issues of their own.
We anticipate that MG/MGOS will facilitate the enhancement of many popular programs and the development of many new programs from diverse communities of computational science and engineering working on the arrangement of spherical objects, including molecules.
\end{sloppypar}
%

\begin{comment}
An immediate example of an important MGOS application is to enhance current MD programs in various directions.
For each time instance of an MD simulation trajectory, geometric features such as voids or channels, along with their geometric properties, can be computed and their evolution over time can be easily traced.
Provided with the dynamic Voronoi diagram \cite{GavrilovaRokne03}, the MGOS can effectively incorporate both hard and soft sphere models in a single algorithmic framework
without maintaining neighbor lists  \cite{DonevEtal05,WeiserEtal98}.
MGOS can also unify both explicit and implicit solvation models because geometric features such as accurate solvent-accessible surfaces can be most efficiently computed using the Voronoi diagram.
Thus, MG is both complementary and supplementary to MD simulation.
It is indeed interesting to discover that one of the earliest computational studies of atomic arrangement was about its geometric characteristics \cite{Bernal59} in a very close connection to molecular dynamics \cite{Bernal62} and was later evolved to the very first use of the ordinary Voronoi diagram of atom centers \cite{BernalFinney67}.
\end{comment}

The challenge remaining is how to identify the set of primitive transformations for geometrization so as to cover as diverse a range of applications and as accurate a set of solutions as possible.
%This can be accomplished
%through close collaboration between geometers and researchers working
%on application problems from diverse domains.
The extensions of MGOS to dynamic situations for moving atoms and to big models such as geometric cell models are also a challenge.
We envision that MG and MGOS together will eventually establish a new paradigm for the computational study of atomic arrangements for both organic and inorganic molecules. MGOS is freely available at http://voronoi.hanyang.ac.kr/software/mgos/.

\section*{Acknowledgment}
This work was supported by the National Research Foundation of Korea (NRF) grant funded by the Korea government (MSIP, MSIT) (Nos. 2017R1A3B1023591, 2016K1A4A3914691).
%

%

%% References


% \nocite{*}
% \bibliographystyle{elsarticle-num}
% \bibliography{sample}






% \end{thebibliography}

\bibliographystyle{elsarticle-num}
\bibliography{biblio}

\clearpage


\appendix
%\title{Supplementary Text:\\ Molecular Geometry and Its Operating System}
\section*{Appendix 1. 300 Test PDB Models}

\begin{table}[htpb]
\caption{
The 300 tested PDB models.
}
\label{tab:pdb 300 data}
\scriptsize
\begin{center}
\begin{tabular}{|c c c c c c c c c c|}
\hline
1AA2 & 1ARL & 1BWW & 1C26 & 1CEX & 1CT4 & 1D2K & 1D4T & 1DC9 & 1DKL \tabularnewline
1DQ0 & 1DQZ & 1E2T & 1EAI & 1EDQ & 1EKG & 1EQP & 1ES9 & 1EUM & 1EY4 \tabularnewline
1EZG & 1F2V & 1F41 & 1F46 & 1F60 & 1FA8 & 1FCQ & 1FHL & 1FQN & 1GCP \tabularnewline
1HM5 & 1I2T & 1I8K & 1IFV & 1ILW & 1IOK & 1IS5 & 1IXV & 1IZ6 & 1IZ9 \tabularnewline
1J27 & 1J2W & 1JEZ & 1JLN & 1JVW & 1JYH & 1K1B & 1K5A & 1KF5 & 1KPK \tabularnewline
1KYF & 1KZ1 & 1L3K & 1L7A & 1L7J & 1LB1 & 1LBW & 1LF1 & 1LHP & 1LN4 \tabularnewline
1LRZ & 1LU9 & 1LW1 & 1LZ1 & 1M0Z & 1M4R & 1M5S & 1M9X & 1MHN & 1MN6 \tabularnewline
1MN8 & 1NKD & 1NLB & 1NR2 & 1NWA & 1ORJ & 1OTV & 1P3C & 1PM4 & 1Q5Z \tabularnewline
1QB5 & 1QKD & 1QP1 & 1QQ1 & 1QXH & 1QZN & 1R0M & 1R0V & 1R1R & 1R1W \tabularnewline
1R29 & 1R2T & 1R3R & 1R4B & 1R5Z & 1R8O & 1RAV & 1RC9 & 1RH9 & 1RL0 \tabularnewline
1RXZ & 1S4F & 1SAU & 1SH5 & 1SNZ & 1SRV & 1SWH & 1SYQ & 1T45 & 1T4Q \tabularnewline
1T5O & 1T6F & 1T7N & 1TM2 & 1TP6 & 1TQG & 1TZQ & 1U07 & 1U3Y & 1UC7 \tabularnewline
1UCS & 1UGQ & 1ULK & 1ULN & 1ULQ & 1VDH & 1VDK & 1VDQ & 1VES & 1VFQ \tabularnewline
1VRX & 1WLG & 1WM3 & 1WU3 & 1WU9 & 1WX0 & 1WYT & 1X13 & 1X25 & 1X7F \tabularnewline
1X91 & 1XG2 & 1XH3 & 1XIX & 1XL9 & 1XMB & 1XMP & 1XN2 & 1XO7 & 1XQO \tabularnewline
1XWG & 1Y0M & 1Y2T & 1Y7Y & 1Y9U & 1YBO & 1YCK & 1YM5 & 1YOY & 1YP5 \tabularnewline
1YPF & 1YVI & 1Z96 & 1ZCF & 1ZEQ & 1ZG4 & 1ZKR & 1ZLB & 1ZLM & 1ZPW \tabularnewline
1ZRS & 1ZS3 & 1ZVT & 1ZWS & 1ZX6 & 1ZYE & 1ZZG & 1ZZK & 2A28 & 2A4V \tabularnewline
2A8F & 2AB0 & 2AHE & 2AQ1 & 2B0J & 2B1K & 2B3M & 2B43 & 2BCM & 2BMA \tabularnewline
2CAR & 2CWC & 2CWL & 2CYG & 2D7T & 2DEP & 2DFU & 2DHH & 2DPO & 2DU7 \tabularnewline
2E3Z & 2ECE & 2EKC & 2EKY & 2EO8 & 2EP5 & 2EQ5 & 2ERF & 2ERW & 2ESK \tabularnewline
2ESN & 2ET6 & 2F51 & 2F6L & 2F82 & 2FBQ & 2FC3 & 2FHZ & 2FIQ & 2FN9 \tabularnewline
2FP8 & 2FTS & 2FU0 & 2G5X & 2G7O & 2G85 & 2GAI & 2GAS & 2GBJ & 2GDG \tabularnewline
2GDN & 2GE7 & 2GFB & 2GG4 & 2GGK & 2GGV & 2GMY & 2GOI & 2GPO & 2GTD \tabularnewline
2GUV & 2H2R & 2H3L & 2H8O & 2HK2 & 2HWX & 2HWZ & 2I1S & 2I3F & 2I49 \tabularnewline
2I6V & 2IC6 & 2IG8 & 2IGD & 2IPB & 2IPR & 2J0N & 2J69 & 2NLS & 2NM0 \tabularnewline
2NVW & 2O37 & 2O70 & 2O7H & 2OBI & 2OEB & 2OL7 & 2ON7 & 2OP6 & 2P19 \tabularnewline
2P2C & 2PET & 2PLQ & 2PLU & 2PN7 & 2QDN & 2QE7 & 2QV3 & 2R57 & 2R6U \tabularnewline
2TMG & 2UX2 & 2V0V & 2VHI & 2VL0 & 2YZ1 & 2Z43 & 2Z5E & 3B7H & 3B8N \tabularnewline
3BB7 & 3BG1 & 3BHS & 3BIP & 3BJV & 3BTU & 3BXY & 3CB4 & 3PVA & 4EUG \tabularnewline
\hline
\end{tabular}
\end{center}
\end{table}

\clearpage
\section*{Appendix 2. APIs of MGOS}

MGOS has several useful API-commands which can be conveniently called from user-created application programs.
Some important current APIs are shown below.
The name of each command explains its task.

\begin{itemize}
\item \textbf{Basic API : Five APIs}
\begin{enumerate}
  \item clear()
  \item load\_atoms( atoms )
  \item preprocess()
  \item get\_all\_atoms()
  \item number\_of\_atoms() \\
\end{enumerate}

\item \textbf{Entity locator API (proximity query I) : Twelve APIs}
\begin{enumerate}
  \item find\_boundary\_atoms\_in\_van\_der\_Waals\_model()
  \item find\_buried\_atoms\_in\_van\_der\_Waals\_model()
  \item find\_first\_order\_neighbor\_atoms\_in\_van\_der\_Waals\_model( atom )
  \item find\_first\_order\_neighbor\_atoms\_in\_van\_der\_Waals\_model( atomArrangement )
  \item find\_second\_order\_neighbor\_atoms\_in\_van\_der\_Waals\_model( atom )
  \item find\_second\_order\_neighbor\_atoms\_in\_van\_der\_Waals\_model( atomArrangement )
  \item find\_boundary\_atoms\_in\_Lee\_Richards\_model( solventProbeRadius )
  \item find\_buried\_atoms\_in\_Lee\_Richards\_model( solventProbeRadius )
  \item find\_first\_order\_neighbor\_atoms\_in\_Lee\_Richards\_model( solventProbeRadius, atom )
  \item find\_first\_order\_neighbor\_atoms\_in\_Lee\_Richards\_model( solventProbeRadius, atomArrangement )
  \item find\_second\_order\_neighbor\_atoms\_in\_Lee\_Richards\_model( solventProbeRadius, atom )
  \item find\_second\_order\_neighbor\_atoms\_in\_Lee\_Richards\_model( solventProbeRadius, atomArrangement ) \\
\end{enumerate}

\item \textbf{Entity verifier API (proximity query II) : Six APIs}
\begin{enumerate}
  \item is\_atom\_on\_boundary\_of\_van\_der\_Waals\_model( atom )
  \item is\_buried\_atom\_van\_der\_Waals\_model( atom )
  \item are\_atoms\_adjacent\_in\_van\_der\_Waals\_model( atom1, atom2 )
  \item is\_atom\_on\_boundary\_of\_Lee\_Richards\_model( solventProbeRadius, atom )
  \item is\_buried\_atom\_of\_Lee\_Richards\_model( solventProbeRadius, atom )
  \item are\_atoms\_adjacent\_in\_Lee\_Richards\_model( solventProbeRadius, atom1, atom2 ) \\
\end{enumerate}

\item \textbf{Entity counter API : Twelve APIs}
\begin{enumerate}
  \item number\_of\_boundary\_atoms\_in\_van\_der\_Waals\_model()
  \item number\_of\_buried\_atoms\_in\_van\_der\_Waals\_model()
  \item number\_of\_first\_order\_neighbor\_atoms\_in\_van\_der\_Waals\_model( atom )
  \item number\_of\_second\_order\_neighbor\_atoms\_in\_van\_der\_Waals\_model( atom )
  \item number\_of\_first\_order\_neighbor\_atoms\_in\_van\_der\_Waals\_model( atomArrangement )
  \item number\_of\_second\_order\_neighbor\_atoms\_in\_van\_der\_Waals\_model( atomArrangement )
  \item number\_of\_boundary\_atoms\_in\_Lee\_Richards\_model( solventProbeRadius )
  \item number\_of\_buried\_atoms\_in\_Lee\_Richards\_model( solventProbeRadius )
  \item number\_of\_first\_order\_neighbor\_atoms\_in\_Lee\_Richards\_model( solventProbeRadius, atom )
  \item number\_of\_second\_order\_neighbor\_atoms\_in\_Lee\_Richards\_model( solventProbeRadius, atom )
  \item number\_of\_first\_order\_neighbor\_atoms\_in\_Lee\_Richards\_model( solventProbeRadius, atomArrangement )
  \item number\_of\_second\_order\_neighbor\_atoms\_in\_Lee\_Richards\_model( solventProbeRadius, atomArrangement ) \\
\end{enumerate}

\item \textbf{Property evaluator API (geometric computation) : Ten APIs}
\begin{enumerate}
  \item compute\_volume\_of\_van\_der\_Waals\_model()
  \item compute\_area\_of\_van\_der\_Waals\_model()
  \item compute\_volume\_and\_area\_of\_van\_der\_Waals\_model()
  \item compute\_voids\_of\_van\_der\_Waals\_model()
  \item compute\_volume\_of\_Lee\_Richards\_model( solventProbeRadius )
  \item compute\_area\_of\_Lee\_Richards\_model(  solventProbeRadius )
  \item compute\_volume\_and\_area\_of\_Lee\_Richards\_model( solventProbeRadius )
  \item compute\_voids\_of\_Lee\_Richards\_model( solventProbeRadius )
  \item compute\_channels( solventProbeRadius, gateSize )
  \item compute\_pockets( ligandSize, solventProbeRadius ) \\
\end{enumerate}
\end{itemize}

\clearpage

\end{document}